\documentclass[aps, pra, twocolumn, superscriptaddress, amsmath, amssymb]{revtex4-2}

\makeatletter
\renewcommand{\fnum@figure}{FIG~\thefigure}
\makeatother

\usepackage{graphicx}
\usepackage{dcolumn}
\usepackage{bm}
\usepackage{mathptmx} 
\usepackage{color}
\usepackage[dvipsnames]{xcolor}
\usepackage[hidelinks,colorlinks=true, linkcolor=WildStrawberry, urlcolor=WildStrawberry, citecolor=Emerald, linktocpage=true, breaklinks=true, bookmarksopen=true]{hyperref}
\usepackage{multirow}
\usepackage[USenglish]{babel}

\begin{document}
	
	\title[Stationary two-qubit entanglement mediated by one-dimensional plasmonic nanoarrays]
	{Stationary two-qubit entanglement mediated by one-dimensional plasmonic nanoarrays} 
	
	\author{Luke C. Ugwuoke}
	\email{ lcugwuoke@gmail.com }
	\affiliation{
		Department of Physics, Stellenbosch University, 
		Private Bag X1, Matieland 7602, South Africa.}
	
	\author{Tjaart P. J. Kr\"{u}ger}
	\affiliation{ 
		Department of Physics, University of Pretoria, 
		Private Bag X20, Hatfield 0028, South Africa.}
	\affiliation{National Institute for Theoretical and Computational Sciences (NITheCS), South Africa.}
	\author{Mark S. Tame}
	\affiliation{
		Department of Physics, Stellenbosch University, 
		Private Bag X1, Matieland 7602, South Africa.}
	\date{\today}
	\hyphenpenalty = 1000
	\begin{abstract}
		Entanglement is one of the key measures of quantum correlations present in nanophotonic systems, with promising applications in quantum optics and beyond. 
		Previous studies have shown that the degree of entanglement between two quantum dot qubits is preserved when a metal nanoparticle is used to mediate the interactions between the qubits.
		In this work, we investigate long-range plasmonic mediation of qubit--qubit entanglement by studying the impact of the number of mediating metal nanoparticles on stationary concurrence.
		Collinear and periodically spaced metal nanoparticles that satisfy the weak-coupling approximation are considered. 
		An effective model that enables the derivation of the mediated interactions within the framework of cavity quantum electrodynamics is employed. 
		Under weak driving at the single particle resonance frequency, the model shows that odd-number arrays are more robust to entanglement decay. 
		We attribute this to strong inter-qubit dissipative coupling as a result of a hybridized dipole plasmon resonating with the driving frequency in odd-number arrays. 
		These arrays can sustain non-vanishing stationary entanglement beyond an inter-qubit spacing of one micron, opening up the possibility of independent spatial optical probing of each quantum dot. 
	\end{abstract}
	
	\maketitle 
	
	\section{Introduction}\label{section1}
	In recent years, quantum correlation metrics, such as discord \cite{Luo2008,Ivano2019}, entanglement \cite{Wootters1998,Dur2000,Dong2022,Kalun2025}, steering \cite{Adesso2015,Yu2022}, and second-order coherence \cite{Ridolfo2010,Martin2011}, have become very useful for investigating non-classical phenomena in quantum systems. Among these metrics, entanglement---where two or more interacting quantum systems are in a shared state---is widely studied as a measure of quantumness in atomic systems \cite{Tanas2004}, nanophotonic systems \cite{Martin2011,He2012,Lee2013,Irfan2024}, and beyond \cite{Wang2009}. The huge interest in entanglement research can be attributed to the many potential applications of the strong quantum correlations possessed by entangled states. For example, entangled photon states have been used to enhance the sensing precision in biosensing \cite{Lee16,Cres12,Tay13,Lee18,Mpofu2022} and image resolution in quantum imaging \cite{Venegas2010,Moodley2024}, as well as to develop reliable cryptographic keys for secure quantum communication channels \cite{Yin2020,Schimp2021,Forbes2024}. Entanglement has also been found to play a significant role in the performance of quantum machines \cite{Wang2009,Brunner2014,Chino2018}, high-speed quantum computations \cite{Josa2003,Zhao2025}, and quantum teleportation networks \cite{Anton1997,Jiang2020}.
	
	Quantum dots (QDs) are nanoscale semiconductor materials with size-dependent electromagnetic properties \cite{Deng2010,Friesen2003,Eriksson2013,Zhou2024}. They have discrete electronic or spin states that can be accessed to create a qubit---a quantum system with two levels \cite{Friesen2003,Eriksson2013,Zhou2024}.
    Two QD qubits may then be entangled either directly or indirectly by interacting with a mediator \cite{Artuso2013}. 
    Several studies have investigated the interaction between a QD and a metal nanoparticle (MNP), where semiclassical effects such as exciton--plasmon coupling \cite{Cox2012,Artuso2013,Ridolfo2010} and quantum effects such as interference-induced two-photon correlation \cite{Ridolfo2010,Waks2010,Zhao2015,Ugwuoke2024} and resonance fluorescence \cite{Ridolfo2010,Zadkov2021} have been reported. Both classical and non-classical effects have also been studied in coupled systems consisting of a QD and a one-dimensional MNP array \cite{August2022,Chan2009,Rink2009}, where the ability of a MNP array to support collective plasmon resonances, as well as confine and transport electromagnetic energy across the nanoparticle chain \cite{Bronger2000,Kelly2003,Maier2003,Zou2006,Park2004,Maier2007,Lee2012,Kitching2013,Dagen2013,Chen2019}, is utilized in enhancing QD absorption \cite{August2022}, Purcell enhancement \cite{Chan2009}, and single-photon generation \cite{Rink2009}. 
	
	Plasmon-mediated bipartite entanglement involves the use of exciton--plasmon coupling between a plasmonic material---the mediator---and a pair of excitonic qubits to generate entangled states under certain conditions \cite{He2012,Hou2014,Nerk2015,Otten2015,Dumit2017,Otten2019,Martin2-2011,Martin2-2011,Li2019}. Due to the inherent ohmic losses present in plasmonic systems, entanglement generation in QD qubits via plasmon-mediated interactions is limited to states that are not maximally-entangled \cite{He2012,Otten2015,Li2019}. 
	Hence, mediated entanglement is actively being researched, with mediators ranging from single MNPs \cite{He2012,Hou2014,Nerk2015,Otten2015} to waveguides \cite{Martin2-2011,Martin2-2011,Li2019,Irfan2024,Qiu2024}. 
    
	Pioneering work in quantum plasmonics showed that polarization-entangled states can be preserved via surface plasmons \cite{Exter2002,Moreno2004}, while recent studies have focused on entanglement generation via plasmonic mediation \cite{Otten2015,Li2019,Kalun2025}, as well as entanglement stabilization \cite{Irfan2024,Vivas2024}.
	With a single MNP, the underlying theory of MNP-mediated entanglement has been well established \cite{He2012,Hou2014}. On the other hand, mediators with waveguiding capabilities aim to sustain qubit--qubit entanglement over large inter-qubit distances where long-range plasmonic coupling due to a single MNP is no longer sufficient to prevent decoherence and the subsequent loss of entanglement \cite{Martin2-2011,Li2019,Qiu2024}. 
	Previous work has also proposed a scheme for long-range two-qubit entanglement generation using a two-arm MNP array, assisted by nanowire waveguides~\cite{Lee2013}. Though efficient, the scheme does not utilize plasmon-mediated inter-qubit interactions, and relies on post-selection, making it hard to realize \cite{Lee2013}. 
	
	In this work, we study the generation of stationary two-qubit entanglement solely via mediated qubit--qubit interactions due to the coupling of the qubits to collective plasmon resonances in a one-dimensional plasmonic nanoarray, and investigate the dependence of the concurrence \cite{Wootters1998,Wootters2001}---a direct entanglement measure for two-qubit systems---on the number, $n$, of MNPs in the array. 
	Previous studies have used the Green's tensor approach to study mediated interactions due to groove-on-metal \cite{Martin2011,Martin2-2011,Li2019} and parallel nanowire \cite{Qiu2024} waveguides. Here, we employ a slightly less cumbersome, but equally reliable approach, based on the effective model previously used in Refs.~\cite{Hou2014,Dumit2017,Vivas2024}, to obtain the mediated interactions due to a MNP array waveguide. 
	
    This paper is organized as follows. The geometry and other parameters of the model, including Hamiltonians, dissipators, and coupling rates of the MNP-MNP and QD-MNP dipole--dipole interactions, are discussed in Sec.~\ref{section2}. 
    In Sec.~\ref{section3}, plasmon-induced terms and mediated coupling rates of the qubits are derived and explained within an effective picture formulation. The effective parameters of the qubits are transformed to the Dicke basis and discussed in Sec.~\ref{section4}, with emphasis on the parameters due to even-$n$ and odd-$n$ arrays. Also, in Sec.~\ref{section4}, the mediated stationary entanglement is discussed and the concurrences due to even and odd $n$-array mediators are compared. A comparison between numerical simulations and the effective model is also provided in Sec.~\ref{section4}, before the conclusion of the paper in Sec.~\ref{section5}. 
    
	\section{Model geometry and parameters}\label{section2}
	\begin{figure} 
	\centering 
	\includegraphics[width=0.475\textwidth]{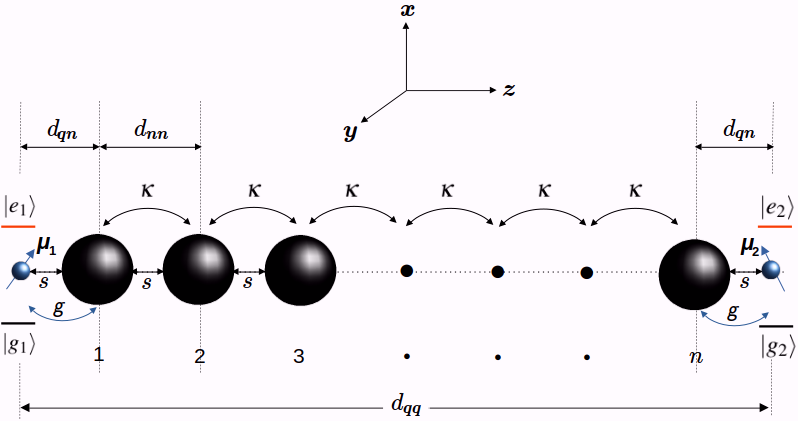}
	\caption{(Color online) Geometry of the QD-NA-QD system. A one-dimensional MNP array is sandwiched by two QD qubits, with dipole moments 
	$\bm{\mu}_{1}$ and $\bm{\mu}_{2}$, respectively. The center-to-center distance between each QD and the nearest MNP is $d_{qn}$. The MNPs in the array are spaced periodically with a center-to-center distance denoted as $d_{nn}$. As the number $n$ of MNPs in the array increases, the qubit--qubit separation, $d_{qq}$, increases. Nearest-neighbor couplings are represented by the QD-MNP dipole--dipole coupling rate $g$ and the MNP-MNP dipole--dipole coupling rate $\kappa$. Each QD qubit has ground and excited states denoted as $|g_{i}\rangle$ and $|e_{i}\rangle$ ($i=1,2$), respectively. 
	}\label{f1}
\end{figure}

The system we consider is shown in Fig.~\ref{f1}. It consists of two QD qubits sandwiching a metal nanoparticle array (NA) at an inter-qubit center-to-center distance denoted by $d_{qq}$. The NA consists of periodically-spaced MNPs with a center-to-center distance $d_{nn}$. Each QD is at a center-to-center distance $d_{qn}$ from the nearest MNP. We denote the surface-to-surface distance between neighboring MNPs and between each QD and the nearest MNP by $s$. The parameters $r_{0}$ and $r$ denote the radii of the QD and MNP, respectively. The distances in Fig.~\ref{f1} are therefore $d_{qn} = r_{0} + s + r$, $d_{nn} = s+2r$, and $d_{qq} = 2(rn+r_{0}) + s(n+1)$, where $n$ is the number of MNPs in the NA. We set $s = r = 30$~nm, satisfying the \emph{point-dipole approximation} \cite{Park2004,Lee2013}: $d_{qn}\ge 2r$ and $d_{nn} \ge 3r$. The QD and MNP can, therefore, be treated as point dipoles with dipole moments $\bm{\mu}_{1} = \bm{\mu}_{2} = \mu_{QD}\hat{e}_{z}$ and $\mu_{\text{MNP}}\hat{e}_{z}$, respectively, where $\hat{e}_{z}$ is a unit vector in the $z$ direction. This approximation allows us to ignore multipolar effects, which usually manifest strongly at coupling distances $s \ll r$ \cite{Zhao2015}. 

Following similar plasmon waveguide-mediated entanglement setups where the illuminating field polarization vector is parallel to the long axis of the waveguide \cite{Martin2-2011,Li2019}, we model the QD-NA-QD system as being driven by a $z$-polarized and time-varying electric field, $\mathbf{E}(t) \approx 2E_{0}\cos(\omega t)\hat{e}_{z}$, propagating in the $x$ direction with a wavevector $\mathbf{k} = k\hat{e}_{x}$. Here, $\omega$ is the driving frequency, $E_{0} = (2I/c \sqrt{\varepsilon_{m}} \varepsilon_{0})^{1/2}$ is the amplitude of the field, with $I$ being the driving field intensity, $\varepsilon_{m}$ the dielectric constant of the host medium, $c$ the speed of light in vacuum, and $\varepsilon_{0}$ the free-space permittivity. We have applied the \emph{long-wavelength approximation} \cite{Park2004,Lee2012}, since the MNP radius with respect to the propagation distance of light in the $x$ direction is in the Rayleigh regime: $r \ll 2\pi c/\omega$. Note that this approximation breaks down if the light field is polarized in the $x$ direction, since this will result in a propagation wavevector in the $z$ direction where $d_{qq} \gtrsim 2\pi c/\omega~\forall~ n>1$. Therefore, we do not consider the $x$-polarized field, since it leads to significant radiation damping due to phase-retardation along the length of the NA. 

As shown in Fig.~\ref{f1}, the QD-MNP and the MNP-MNP coupling rates are denoted as $g$ and $\kappa$, respectively. The above approximations allow us to consider nearest-neighbor dipole--dipole interactions only, since these coupling strengths dominate the QD-MNP and MNP-MNP interactions at such length scales \cite{Lee2013, Lee2012}. In addition, they are small compared to the driving frequency, satisfying the \emph{weak-coupling approximation} \cite{Lee2013}: $g, \kappa \ll \omega = \omega_{0}$, where  $\omega_{0}$ denotes the localized surface plasmon resonance (LSPR) of a single MNP in the NA. Refs.~\cite{Ridolfo2010,Waks2010,Cox2012, Zhao2015,Kim14} have shown how $g$ is derived. A similar approach can be applied to an MNP homodimer (see Sec. 1 of the Supplemental Material (SM) \cite{Supp}) to obtain $\kappa$ and $\mu_{\text{MNP}}$. Hence, we have the following equations: 
\begin{subequations}
\begin{align}
g & = \frac{s_{z}\mu_{QD}}{d^{3}_{qn}}\sqrt{\frac{3r^{3}\eta}{4\pi\varepsilon_{0}\hbar}}, \label{e1a}\\
\kappa & = 3s_{z}\varepsilon_{m}\eta\left(\frac{r}{d_{nn}} \right)^{3}, \label{e1b} \\
\mu_{\text{MNP}} & = 2\varepsilon_{m}\sqrt{3\pi\varepsilon_{0}\hbar\eta r^{3}}, \label{e1c}
\end{align}
\end{subequations}
where $s_{z} = +2$ is a dipole orientation factor corresponding to longitudinal coupling of the dipole moments involved in each coupling rate, $\mu_{QD} = er_{0}$, with $e$ being the electronic charge, $\eta \approx \omega_{0}/2(\varepsilon_{\infty} +2\varepsilon_{m})$, $\omega_{0} \approx \omega_{p}/\sqrt{\varepsilon_{\infty} +2\varepsilon_{m} }$, with $\varepsilon_{\infty}$ and $\omega_{p}$ being the high-frequency permittivity
and bulk plasma frequency of the metal, respectively, 
and $\hbar$ is the Dirac constant. 

The coherent part of the system's dynamics is described by the following
Jaynes-Cummings type Hamiltonian:
\begin{equation}
H = H_{QD}  + H_{NA} + H_{int} + H_{ext}, \label{e2}
\end{equation} 
where $H_{QD}$ and $H_{NA}$ are the free Hamiltonians of the QDs and NA, 
respectively, $H_{int}$ is the interaction Hamiltonian of the QD-NA-QD system, and $H_{ext}$ is the Hamiltonian representing the interaction of the system with the external electric field.
Based on previous works \cite{Lee2012,Lee2013,Hou2014}, the components of the Hamiltonian in Eq. \eqref{e2}, in a frame rotating with the driving frequency, can be written as
\begin{subequations}
\begin{align}
H_{QD} &  = \hbar\sum_{i = 1}^{2} \Delta\omega_{i}\sigma^{\dagger}_{i}\sigma_{i}, \label{e3a}\\
H_{NA} &  = \hbar\sum_{m = 1}^{n} \Delta\omega_{0}a^{\dagger}_{m}a_{m} 
-\hbar\sum_{m,v|m<v}\kappa_{mv}\left(a^{\dagger}_{m}a_{v} +  a_{m}a_{v}^{\dagger} \right), \label{e3b}\\
H_{int} &  = -\hbar \sum_{m = 1}^{n}\sum_{i = 1}^{2}g_{im}(\sigma^{\dagger}_{i}a_{m} + \sigma_{i}a^{\dagger}_{m}), \label{e3c} \\
H_{ext}  &  = -\hbar\sum_{i = 1}^{2} \Lambda_{i}(\sigma^{\dagger}_{i} + \sigma_{i}) - \hbar\sum_{m = 1}^{n} \Omega_{m}(a^{\dagger}_{m} + a_{m}), \label{e3d} 
\end{align}
\end{subequations}
where $\kappa_{mv} = \kappa$ for $v\in $ nearest-neighbor$(m)$ and zero otherwise, $g_{11} = g_{2n} = g$, 
$g_{1(m>1)} = g_{2(m<n)} = 0$ (i.e., only nearest-neighbor QD-MNP couplings are non-zero), 
$\Delta\omega_{i} = \omega_{i} - \omega$ is the detuning of the transition frequency of each QD from the driving frequency, $\sigma^{\dagger}_{i}(\sigma_{i})$ is the raising: $|e_{i}\rangle\langle g_{i}|$ (lowering: $|g_{i}\rangle\langle e_{i}|$) operator of each QD, $\Lambda_{i} = E_{0}\mu_{i}/\hbar$ is the excitation rate of each QD, 
$\Delta\omega_{0} = \omega_{0} - \omega$ is the detuning of the LSPR of each MNP from the driving frequency, $a^{\dagger}_{m}(a_{m})$ is the plasmon creation (annihilation) operator of each MNP, and $\Omega_{m} = E_{0}\mu_{\text{MNP}}/\hbar$ is the excitation rate of each MNP. 

At optical frequencies, the Liouvillian describing the dissipative part of the evolution of the reduced density matrix $\rho$ of the system due to its interaction with the environment, within the Markov approximation \cite{Waks2010,Lee2012}, is given by 
\begin{equation} \label{e4}
L\rho = \frac{1}{2}\sum_{i = 1}^{2}\gamma_{i}D_{i}[\sigma_{i},\rho] + 
 \frac{1}{2}\sum_{m = 1}^{n}\gamma_{0}D_{m}[a_{m},\rho],
\end{equation}
where $D_{i}[\sigma_{i},\rho] = 2\sigma_{i}\rho\sigma^{\dagger}_{i} - \{\sigma^{\dagger}_{i}\sigma_{i},\rho \}$ and $D_{m}[a_{m},\rho] = 2a_{m}\rho a^{\dagger}_{m} - \{a^{\dagger}_{m}a_{m},\rho \}$ are the dissipators consisting of contributions from the spontaneous emission rate $\gamma_{i}$ of the $i$-th QD qubit and plasmon damping rate $\gamma_{0}$ of the $m$-th MNP, respectively. The plasmon damping rate, 
$\gamma_{0} = \gamma_{nr} + \gamma_{r}$, is due to both non-radiative damping, $\gamma_{nr} = \gamma_{p}(1 + (\gamma_{p}/\omega_{0})^{2})$, with $\gamma_{p}$ being the damping rate of the free electrons of the metal, and radiative damping, $\gamma_{r} = \mu_{\text{MNP}}^{2}\sqrt{\varepsilon_{m}}\omega^{3}_{0}/3\pi\varepsilon_{0}\hbar c^{3}$, of the dipole plasmon \cite{Waks2010,Ugwuoke2024}. 

\section{Effective model}\label{section3}
In addition to the weak-coupling approximation, the effective approach
used in Refs.~\cite{Hou2014,Dumit2017} and adopted here also requires the system to be driven weakly. 
In the weak-field limit, $\Omega_{m} \ll \gamma_{0}$ (since $\Omega_{m} \gg \Lambda_{i}$ and $\gamma_{0} \gg \gamma_{i}$), the joint expectation values of the MNP and QD operators appearing in the equations of motion (EOM) can be factored \cite{Waks2010}. 
To obtain the EOM, the evolution of the reduced density matrix of the system is described by the Lindblad equation \cite{Waks2010,Carmichael2013,Gard04}: 
\begin{equation}\label{e5}
\dot{\rho} = \frac{i}{\hbar}[\rho,H] + L\rho.
\end{equation}
Using $tr( a_{m}\dot{\rho}  ) = \langle \dot{a}_{m} \rangle$ and  $tr( \sigma_{i}\dot{\rho}  ) = \langle \dot{\sigma}_{i} \rangle$,
we obtain the following EOM: 
\begin{subequations}
\begin{align}
\langle \dot{a}_{m} \rangle & = -\delta\langle a_{m} \rangle +i\Big(\Omega_{m} + \sum_{v}\kappa_{mv}\langle a_{v} \rangle \Big) \nonumber\\
&+\langle \sigma_{1} \rangle g_{1m} + \langle \sigma_{2} \rangle g_{2m}, \label{e6a}\\
\langle \dot{\sigma}_{i} \rangle & = -\delta_{i}\langle \sigma_{i} \rangle -i\Big(\Lambda_{i} + \sum_{m = 1}^{n}g_{im} \langle a_{m} \rangle \Big)\langle \sigma^{i}_{z} \rangle, \label{e6b}
\end{align}
\end{subequations}
where $\delta = i\Delta\omega_{0} + \gamma_{0}/2, \delta_{i} = i\Delta\omega_{i} + \gamma_{i}/2$, 
$\langle \sigma^{i}_{z} \rangle$ is the expectation value of the population difference operator, $\sigma^{i}_{z} = 2\sigma^{\dagger}_{i}\sigma_{i} - 1$, of each QD qubit, 
the expectation values of the MNP and QD fluctuation operators are zero, and joint expectation values have been factored. 
Since $\gamma_{0} \gg \gamma_{i}$, the MNP dynamics can be treated as stationary with respect to the dynamics of the QD qubits \cite{Hou2014,Dumit2017,Chino2018}. This enables the expectation value of the MNP annihilation operator to be eliminated adiabatically from the qubit dynamics using the stationary solution of Eq.~\eqref{e6a}: 
\begin{equation}\label{e7}
\langle a_{m} (t\rightarrow\infty) \rangle  = \frac{i}{\delta}\Big(\tilde{\Omega}_{m} +  \tilde{g}_{1m}\langle \sigma_{1} \rangle + \tilde{g}_{2m}\langle \sigma_{2} \rangle \Big),
\end{equation}
where $\tilde{\Omega}_{m} = \sum_{v}K_{mv}\Omega_{v}$ is the effective excitation rate of each MNP due to the nearest-neighbor MNP-MNP interactions, $
\tilde{g}_{1m} = \sum_{v}g_{1v}K_{mv}$ and $\tilde{g}_{2m} = \sum_{v}g_{2v}K_{mv}$ are the effective QD-MNP coupling rate of each QD, respectively, and $K_{mv}$ are the elements of the inverse of the coupling matrix $A^{(n)}$---an $n\times n$ square matrix with diagonal entries $A^{(n)}_{ii} = 1$ and off-diagonal entries $A^{(n)}_{ij} = -i\kappa/\delta$ (for nearest-neighbor interactions) and $A^{(n)}_{ij} = 0$ (otherwise). Substituting 
Eq.~\eqref{e7} into Eq.~\eqref{e6b} for $\langle a_{m} \rangle$ leads to
\begin{equation}
\langle \dot{\sigma}_{i} \rangle = -\tilde{\delta}_{i}\langle \sigma_{i} \rangle - i\tilde{\Lambda}_{i}\langle \sigma^{i}_{z} \rangle \\
 -\Big(i\tilde{G}_{ij} - \frac{1}{2}\tilde{\Gamma}_{ij}\Big) \langle\sigma^{i}_{z} \rangle \langle \sigma_{j} \rangle, 
\end{equation}
with $\tilde{\delta}_{i} = (i\Delta\tilde{\omega}_{i} + \frac{1}{2}\tilde{\gamma}_{i})$, 
where the plasmon-induced effects
\begin{subequations}
\begin{align}
\Delta\tilde{\omega}_{i} & = \Delta\omega_{i} - \sum_{m=1}^{n} g_{im}
\frac{ V_{im} }{|\delta|^{2}}, \label{e9a}\\
\tilde{\gamma}_{i} & = \gamma_{i} + \sum_{m=1}^{n} g_{im}
\frac{ 2U_{im} }{|\delta|^{2}}, \label{e9b}\\
\tilde{\Lambda}_{i} & =  \Lambda_{i} + \sum_{m=1}^{n} g_{im}
\frac{i \tilde{\Omega}_{ m} }{\delta}, \label{e9c}
\end{align}
\end{subequations}
are the exciton shift, Purcell enhancement, and excitation rate enhancement, respectively, of the QD qubits, and 
\begin{equation} \label{e10}
\tilde{G}_{ij} = \sum_{m=1}^{n} g_{im}
\frac{ V_{jm} }{|\delta|^{2}}, ~~~
\tilde{\Gamma}_{ij} = \sum_{m=1}^{n} g_{im}
\frac{ 2U_{jm} }{|\delta|^{2}}, 
\end{equation}
are the plasmon-mediated interactions, with matrix elements 
$V_{jm} = \Delta\omega_{m}\Re[ \tilde{g}_{jm} ] - 
\frac{1}{2}\gamma_{0}\Im[ \tilde{g}_{jm} ]$ and $U_{jm} = \Delta\omega_{m}\Im[ \tilde{g}_{jm} ] + \frac{1}{2}\gamma_{0}\Re[ \tilde{g}_{jm} ]$. 
In Eq.~\eqref{e10}, $\tilde{G}_{ij}$ is the coherent coupling rate, responsible for the unitary evolution of the qubits, while their non-unitary dynamics is due to the non-radiative, dissipative coupling rate $\tilde{\Gamma}_{ij}$ and the effective spontaneous emission rate $\tilde{\gamma}_{i}$ given by Eq.~\eqref{e9b}.
Thus, the MNPs act as mediators by providing the above mediated interactions that act in place of the intrinsic qubit-qubit interactions \cite{Tanas2004,Artuso2013}.
Based on Eqs.~\eqref{e9a}-\eqref{e9c} and \eqref{e10} and the effective model in 
Refs.~\cite{Hou2014,Martin2-2011}, we can write the effective Hamiltonian and Liouvillian of the NA-coupled QD qubits as follows: 
\begin{subequations}
\begin{align}
\tilde{H} & = \hbar\sum_{i=1}^{2}\Big(\Delta\tilde{\omega}_{i}\sigma_{i}^{\dagger}\sigma_{i} - (\tilde{\Lambda}_{i}\sigma_{i}^{\dagger} + \tilde{\Lambda}_{i}^{*}\sigma_{i} )  \Big) \nonumber \\
&- \hbar\sum_{i,j = 1}^{2}G_{ij}(1-\delta_{ij})\sigma_{i}^{\dagger}\sigma_{j}, \label{e11a} \\
\tilde{L}\rho & =  \frac{1}{2}\sum_{i,j=1}^{2}\tilde{\Gamma}_{ij}(2\sigma_{i}\rho\sigma^{\dagger}_{j} - \{\sigma^{\dagger}_{i}\sigma_{j},\rho \}), \label{e11b}
\end{align}
\end{subequations}
with $\tilde{\Gamma}_{ii} = \tilde{\gamma}_{i}$. 

\begin{figure} 
	\centering 
	\includegraphics[width=0.40\textwidth]{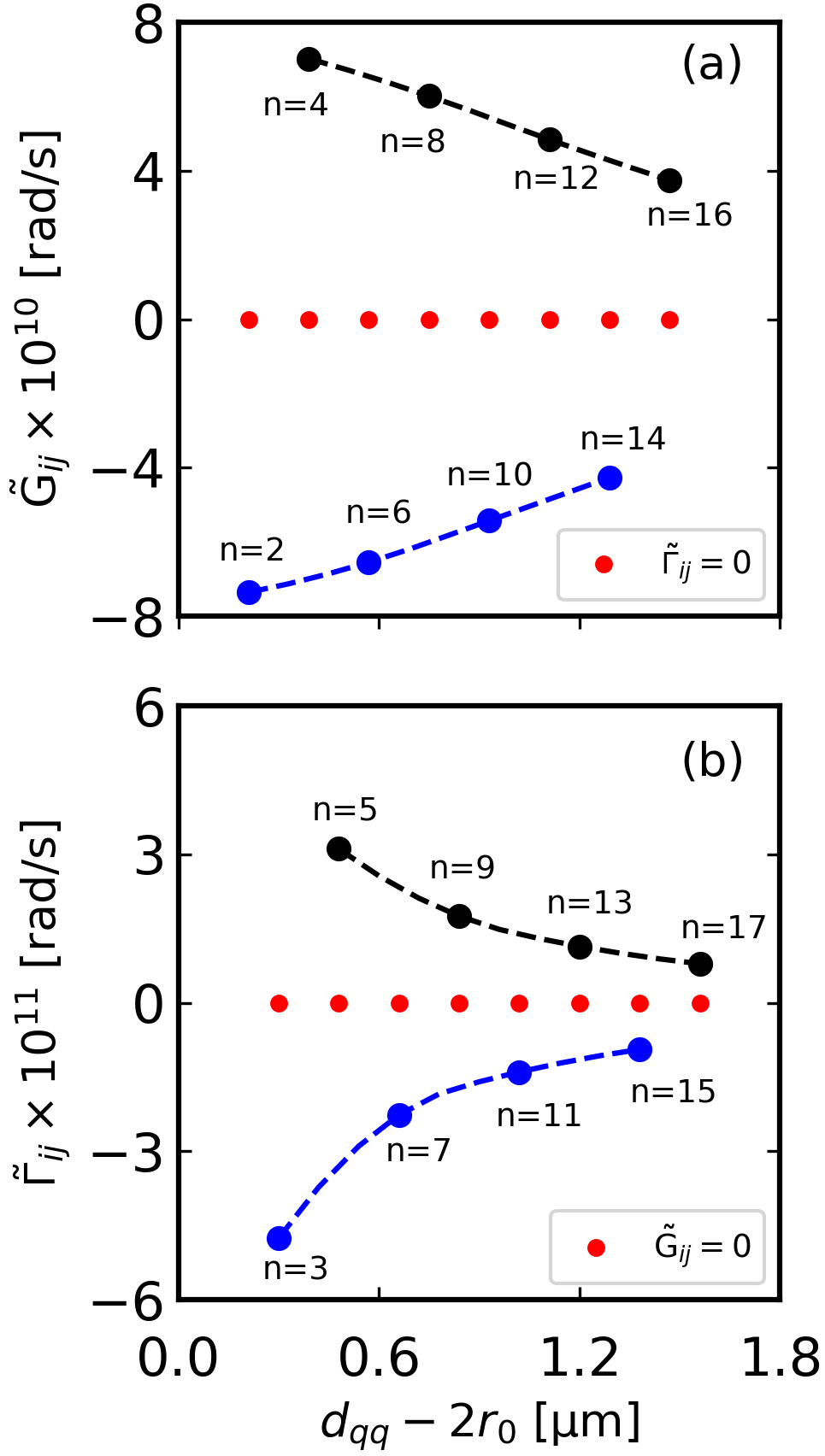}
	\caption{(Color online) Dependence of the coherent coupling, $\tilde{G}_{ij}$ and the dissipative coupling, $\tilde{\Gamma}_{ij}$ ($i = 1, j = 2$), on the inter-qubit separation $d_{qq}-2r_{0}$ for (a) even-$n$ arrays: $\tilde{G}_{ij} \neq 0$, $\tilde{\Gamma}_{ij} = 0$ and (b) odd-$n$ arrays: $\tilde{G}_{ij} = 0$, $\tilde{\Gamma}_{ij} \neq 0$ , at $\omega = \omega_{0} \approx 2\pi c/(480$~nm). Dashed curves: polynomial fits. Solid points: data. 
		}\label{f2}
\end{figure}

The NA investigated is that of silver nanoparticles (AgNPs). We assume that each AgNP in the NA is surrounded by a medium of dielectric constant $\epsilon_{m} \approx 2.98$. This leads to a single particle resonance frequency $\omega_{0} \approx 2\pi c/(480$~nm), obtained with the following Drude parameters: $\omega_{p} \approx 8.5472$~eV, $\varepsilon_{\infty} \approx 5$ and $\gamma_{p} = 0.018$~eV. Here, $\omega_{p}$ and $\varepsilon_{\infty}$ have been adjusted to reproduce typical LSPR values of an AgNP on silicon-based substrates \cite{Zadeh2017}. We consider two similar QD qubits, each with radius $r_{0}  = 2$~nm (corresponding to $\mu_{QD} \sim 96$~Debye), spontaneous emission rate $\gamma_{i} = 2\pi \times10^{8}$~rad/s, and a transition frequency $\omega_{i} \approx \omega_{0}$, such as tunable cadmium-based core-shell quantum dots \cite{Deng2010}. 

As shown in Fig.~\ref{f2}, the plasmon-mediated couplings in 
Eq.~\eqref{e10} are affected by the number, $n$, of MNPs in the NA, at the driving frequency $\omega = \omega_{0}$ of the single MNP LSPR. 
Though the coherent coupling rate, $\tilde{G}_{ij}$, and the dissipative coupling rate, $\tilde{\Gamma}_{ij}$, are independent of the driving field intensity, they oscillate as the driving frequency $\omega$ is varied, reaching crests or troughs, depending on the value of $n$ (see Fig.~S2 of the SM \cite{Supp}). These mediated coupling rates have positive values at the crests and negative values at the troughs. The values displayed in Fig.~\ref{f2} are for $\omega = \omega_{0}$, since they are the optimal inter-qubit coupling rates when the QD-NA-QD system is driven resonantly at the single MNP LSPR. 

The polynomial fits in Fig.~\ref{f2} reveal the dependence of the NA-mediated interactions---$\tilde{G}_{ij}$ and $\tilde{\Gamma}_{ij}$---on the qubit-qubit separation. The fits show that the MNP array waveguide leads to mediated interactions with a nonlinear dependence on the inter-qubit separation, in constrast with certain plasmon waveguides \cite{Li2019}, 
where the mediated interactions have a linear dependence on the inter-qubit distance. Here, both $\tilde{G}_{ij}$ and $\tilde{\Gamma}_{ij}$ are quadratic in $d_{qq}-2r_{0}$, but the nonlinearity is more pronounced in $\tilde{\Gamma}_{ij}$, as shown in Fig.~\ref{f2} (b). 
This is likely due to differences in the number of hybrid plasmon resonances as well as other factors, such as plasmon damping and qubit-NA interactions at the driving frequency, $\omega = \omega_{0}$. 

\begin{figure*} 
	\centering 
	\includegraphics[width=0.33\textwidth]{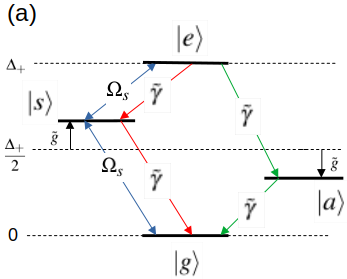}~~~
	\includegraphics[width=0.32\textwidth]{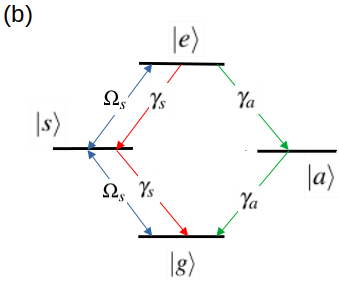}
	\includegraphics[width=0.32\textwidth]{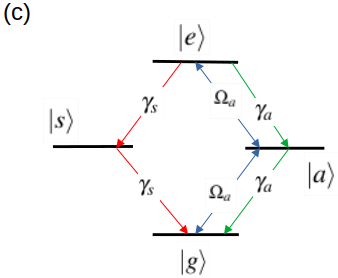}
	\caption{(Color online) Energy-level diagrams showing the transition rates at the driving frequency $\omega = \omega_{0}$ between the Dicke states of the two NA-coupled qubits for the (a) even-$n$ array with symmetric excitation rate, $\Omega_{s}$, (b) odd-$n$ array with symmetric excitation rate, $\Omega_{s}$, and (c) odd-$n$ array with antisymmetric excitation rate, $\Omega_{a}$. 
	}\label{f3}
\end{figure*}

When $n$ is even, Fig.~\ref{f2} (a) shows that the qubits can only interact via coherent coupling (since $\tilde{\Gamma}_{ij} = 0$). This is because the oscillatory behavior of the dissipative coupling in an even-$n$ array is out-of-phase with the driving field at $\omega = \omega_{0}$, while that of the coherent coupling is in-phase with the field. The converse is the case for odd-$n$ arrays, i.e., the qubits can only interact via dissipative coupling when $n$ is odd (since $\tilde{G}_{ij} = 0$), as shown in Fig.~\ref{f2} (b). 

For certain values of $n$, the inter-qubit couplings can either attain their crests or troughs at $\omega = \omega_{0}$. This leads to different $n$ array sequences, each with a common difference of $4$, as shown in Fig.~\ref{f2}. 
The constant common difference arises due to the behavior of the mediated couplings at $\omega = \omega_{0}$. For instance, a trough in $\tilde{G}_{ij}$ at $n=2$ is followed by a trough in $\tilde{\Gamma}_{ij}$ at $n = 3$, a crest in $\tilde{G}_{ij}$ at $n=4$, then a crest in $\tilde{\Gamma}_{ij}$ at $n=5$, before the next trough in $\tilde{G}_{ij}$ at $n=6$. 
The dissipative coupling rate (Fig.~\ref{f2} (b)) for odd-$n$ arrays is about an order of magnitude greater than the coherent coupling rate (Fig.~\ref{f2} (a)) for even-$n$ arrays. This is because, unlike even-$n$ arrays, where the hybrid modes are off-resonant, odd-$n$ arrays support a hybridized LSPR at $\omega = \omega_{0}$, leading to enhanced interaction due to resonant coupling. 
As $n$ increases, the inter-qubit distance increases via $d_{qq}$, causing both $|\tilde{\Gamma}_{ij}|$ and $|\tilde{G}_{ij}|$ to decrease due to their dipole character. 
As we discuss in the next section, the positive and negative values of $\tilde{\Gamma}_{ij}$ and $\tilde{G}_{ij}$ enable optimal control of the energy levels and transition rates involved in entanglement mediation with the NA. 

\section{Mediated entanglement}\label{section4}
\subsection{Transition rates}
It is often useful to discuss the transition rates between the energy levels of the two-qubit states involved in bipartite entanglement in the Dicke basis \cite{Tanas2004,Hou2014,Chino2018}: $|g\rangle  = |0\rangle, |s\rangle  = \frac{1}{\sqrt{2}}(|1 \rangle +|2 \rangle), 
|a\rangle = \frac{1}{\sqrt{2}}(|1 \rangle - |2\rangle), |e\rangle  = |3\rangle$, where the qubit--qubit Hamiltonian given by Eq. \eqref{e11a} is diagonal, 
$|g\rangle$ and $|e\rangle$ are the ground and excited states of the two QD qubits, while $|s\rangle$ and $|a\rangle$ are Bell states, referred to as symmetric and antisymmetric states of the qubits, respectively. Here, explicitly, the states  $|0\rangle  = |g_{1}\rangle \otimes |g_{2}\rangle$, $|1\rangle  = |e_{1}\rangle \otimes |g_{2}\rangle$, 
$|2\rangle  = |g_{1}\rangle \otimes |e_{2}\rangle$, and $|3\rangle  = |e_{1}\rangle \otimes |e_{2}\rangle$ are in the so-called computational basis~\cite{Tanas2004}. 
In the following, we transform Eqs. \eqref{e11a} and \eqref{e11b} to the Dicke basis using the collective operators: $S = (\sigma_{1} + \sigma_{2})/\sqrt{2}$ and $A = (\sigma_{1} - \sigma_{2})/\sqrt{2}$ ~\cite{Chino2018}, denoting the symmetric and 
antisymmetric operators, respectively, to obtain: 
\begin{subequations}
\begin{align}
\tilde{H} & = \hbar(E_{+}S^{\dagger}S + E_{-}A^{\dagger}A + \Delta_{-}(S^{\dagger}A + A^{\dagger}S)) \nonumber \\
          & -\hbar(S^{\dagger}\Omega_{s} + A^{\dagger}\Omega_{a} + S\Omega^{*}_{s} + A\Omega^{*}_{a}), \label{e12a} \\
\tilde{L}\rho  & = \frac{\gamma_{s}}{2}(2S\rho S^{\dagger} - \{S^{\dagger}S,\rho \}) + \frac{\gamma_{a}}{2}(2A\rho A^{\dagger} - \{A^{\dagger}A,\rho \}). \label{e12b}
\end{align}
\end{subequations}
In the above, $E_{\pm} = \Big(\frac{\Delta_{+}}{2} \pm \tilde{g} \Big)$, $\tilde{g} = \tilde{G}_{12} = \tilde{G}_{21}$, 
$\Delta_{+} = (\tilde{\Delta}\omega_{1}+\tilde{\Delta}\omega_{2})/2$, 
$\Delta_{-} = (\tilde{\Delta}\omega_{1}-\tilde{\Delta}\omega_{2})/2$,
$\Omega_{s} = (\tilde{\Lambda}_{1}+\tilde{\Lambda}_{2})/\sqrt{2}$, $\Omega_{a} = (\tilde{\Lambda}_{1}-\tilde{\Lambda}_{2})/\sqrt{2}$, 
$\gamma_{s} = \tilde{\gamma} + \tilde{\Gamma}_{12}$, $\gamma_{a} = \tilde{\gamma} - \tilde{\Gamma}_{12}$, $\tilde{\gamma} = \tilde{\gamma}_{1} = \tilde{\gamma}_{2}$, $\tilde{\Gamma}_{12} = \tilde{\Gamma}_{21}$, 
 $S^{\dagger}S = |e\rangle \langle e| + |s\rangle \langle s|$, $A^{\dagger}A = |e\rangle \langle e| + |a\rangle \langle a|$, $S = |g\rangle \langle s| + |s\rangle \langle e|$, and 
 $ A = |g\rangle \langle a| - |a\rangle \langle e|$.

\begin{figure*} 
	\centering 
	\includegraphics[width=0.365\textwidth]{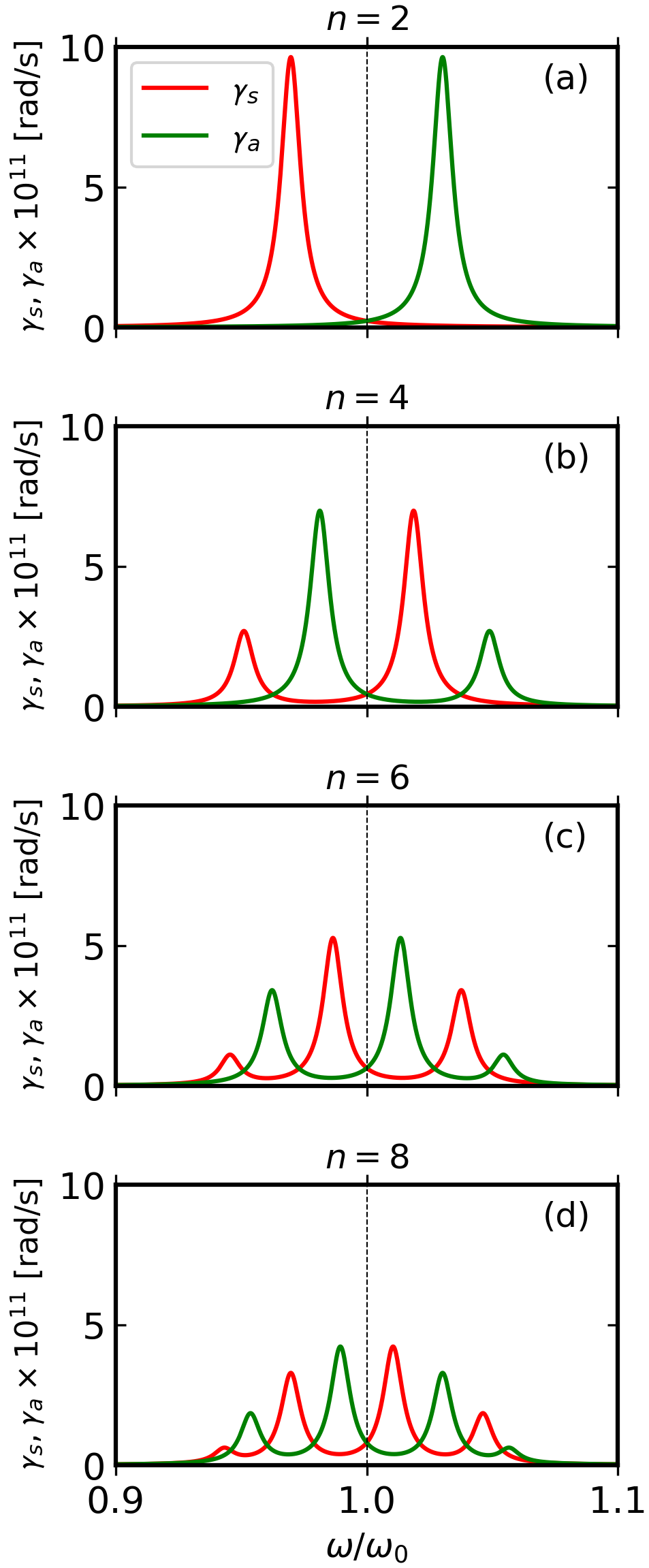}~
	\includegraphics[width=0.330\textwidth]{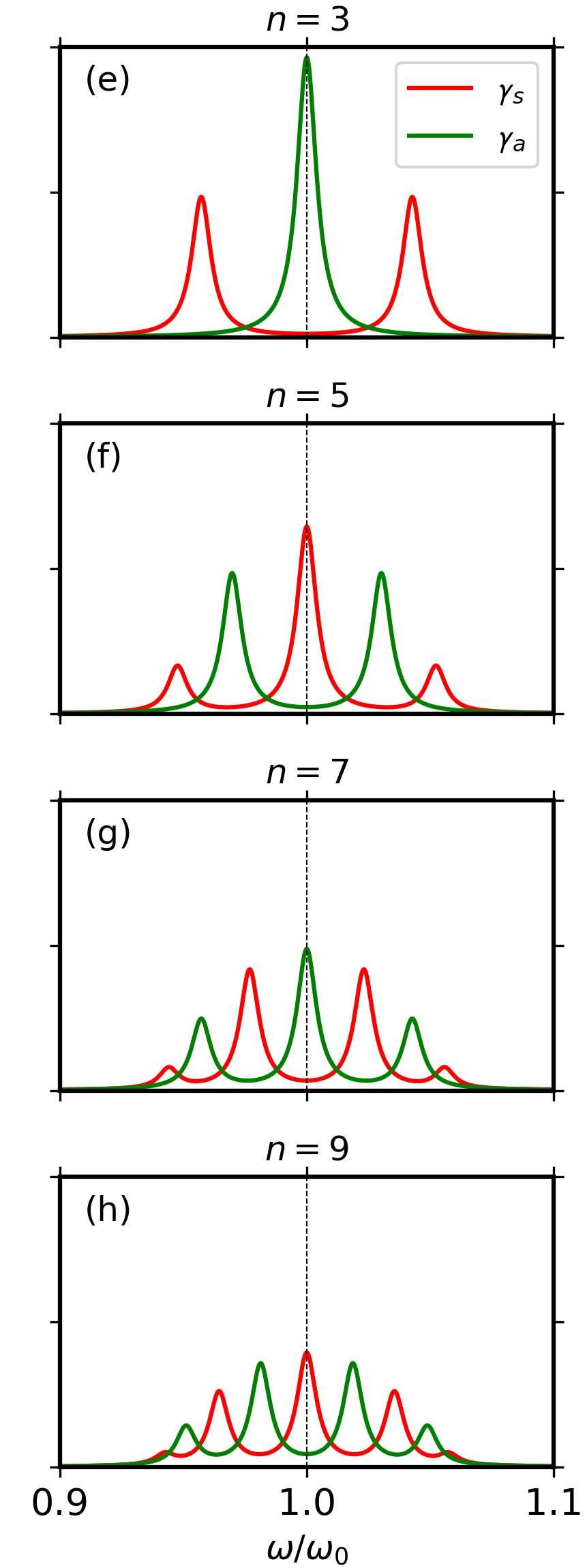}
	\caption{(Color online) Symmetric and antisymmetric decay rates, $\gamma_{s}$ and $\gamma_{a}$, as a function of the driving frequency $\omega$ for even-$n$ arrays: $n = 2, 4, 6, 8$ and odd-$n$ arrays: $n = 3, 5, 7, 9$. The dashed vertical line in each panel represents the position of the single particle LSPR,  $\omega_{0}$. 
	}\label{f4}
\end{figure*}

Based on Eqs.~\eqref{e12a} and \eqref{e12b}, the energy-level diagrams showing the transition rates between the four Dicke states are constructed as shown in Fig.~\ref{f3}, for even-$n$ and odd-$n$ NA mediators. When $n$ is even (Fig. \ref{f3} (a)), $\tilde{G}_{ij} \neq 0$ and $\tilde{\Gamma}_{ij} = 0$ at $\omega = \omega_{0}$, so that $E_{+} \neq E_{-}$ and $\gamma_{s}  = \gamma_{a} = \tilde{\gamma}$, respectively, i.e., the energy levels of the symmetric and antisymetric states split while their decay rates remain the same. 
Figs.~\ref{f4} (a)-(d) further show why this is the case for the decay rates when $n$ is even. Here, the hybrid plasmon modes of the NA split into $n$ modes, which are symmetric around the driving frequency. 
The modes contribute to the symmetric and antisymmetric decay rates, $\gamma_{s}$ and $\gamma_{a}$, of the coupled qubits. For instance, when $n=2$ 
(Fig.~\ref{f4} (a)), the hybridized plasmon resonance splits into two single modes, each contributing to either one of the decay rates---the higher-energy mode ($\omega>\omega_{0}$) contributes to $\gamma_{s}$ (red curve), while $\gamma_{a}$ (green curve) is due to a lower-energy mode ($\omega<\omega_{0}$). As $n$ increases, the number of higher- and lower-energy modes contributing to the decay rates increases. This is shown in Figs.~\ref{f4} (b)-(d). For example, when $n=4$ (Fig.~\ref{f4} (b)), the hybridized modes split into two dimer modes, each contributing to the decay rates. 

However, as shown in Figs.~\ref{f4} (a)-(d), the mode splitting is such that the two decay rates coincide at the driving frequency, $\omega = \omega_{0}$. Hence, when $n$ is even, the symmetric and antisymmetric states, $|s\rangle$ and $|a\rangle$, will decay to the ground state, $|g\rangle$, at the same rate---the effective spontaneous emission rate, $\tilde{\gamma}$, of the qubit---while the bi-excited state, $|e\rangle$, will decay to $|g\rangle$ at the rate $2\tilde{\gamma}$, as illustrated in Fig.~\ref{f3} (a). 
Figs.~\ref{f4} (a)-(d) show that as $n$ increases, $\tilde{\gamma}$ increases at $\omega = \omega_{0}$, making the system more lossy and the even-$n$ array less efficient as an entanglement mediator, through an enhanced decay to the ground state via $\tilde{\gamma}$. 
In addition, since $\tilde{g} = \tilde{G}_{ij}$, the energy level of the symmetric state, shown in Fig.~\ref{f3} (a), undergoes either an upward or a downward shift while that of the antisymmetric state shifts downward or upward, depending on the sequence of the even-$n$ array mediating the qubit--qubit interaction. 

To optimize entanglement mediation with the even-$n$ array, 
the qubits are driven symmetrically and detuned symmetrically from the single-particle LSPR, $\omega_{0}$. Symmetric driving via $\tilde{\Lambda}_{1} = \tilde{\Lambda}_{2}$, results in $\Omega_{s} \neq 0$ and $\Omega_{a} = 0$, giving the energy-level diagram shown in Fig.~\ref{f3} (a). 
This allows the symmetric state population, $\rho_{ss}$, to contribute most to the degree of stationary entanglement when $n$ is even. 
The qubit detunings are chosen such that when $\tilde{g}$ is positive, the amount of detuning, $\Delta_{1} = \Delta_{2} = \Delta$, for each qubit (via $\omega_{1} = \omega_{0} + \Delta_{1}$ and $\omega_{2} = \omega_{0} + \Delta_{2}$), is also positive, and vice versa, leading to 
$\Delta_{+} = \pm2\Delta, \Delta_{-} = 0$, and $E_{\pm}  = +\Delta \pm |\tilde{g}|$ for positive $\tilde{g}$ ($n = 4, 8, 12, ...$) and $E_{\pm}  = -\Delta \mp |\tilde{g}|$ for negative $\tilde{g}$ ($n = 2, 6, 10, ...$).
The term involving $\Delta_{-}$, which couples $|s\rangle$ to $|a\rangle$ is therefore not present in 
Eq. \eqref{e12a}, further facilitating the build up of $\rho_{ss}$. 
As we explain in more detail in the next section, the value of $\Delta$ is chosen such that it has a similar order of magnitude as $\tilde{g}$, otherwise the role of $\tilde{g}$ in tuning the energy levels of $|e\rangle$, $|s\rangle$ and $|a\rangle$ becomes insignificant. 

When $n$ is odd, $\tilde{G}_{ij} = 0$ at $\omega = \omega_{0}$, hence there is no shift in the energy levels of the symmetric and antisymmetric states, i.e., $E_{+} = E_{-}$, as illustrated in Figs.~\ref{f3} (b) and (c). However, as shown in Figs.~\ref{f4} (e)-(h), the mode splitting behavior of the hybridized plasmon modes in the symmetric and antisymmetric decay rates of the qubits is such that, while the splitting still consists of $n$ plasmon resonances, the splitted modes contribute unequally to the symmetric and antisymmetric decay rates of the odd $n$-array coupled qubits.
For instance, when $n=3$ (Fig.~\ref{f4} (e)), a hybrid single mode at $\omega = \omega_{0}$ is responsible for the antisymmetric decay rate (green curve), while the symmetric decay rate is due to a hybrid dimer mode (red curve), consisting of both higher- and lower-energy modes. This is because when $n=3$,  $\tilde{\Gamma}_{ij}$ is negative, which leads to $\gamma_{a} > \gamma_{s}$ at $\omega = \omega_{0}$, corresponding to the LSPR of the hybrid single mode. 
However, when $n=5$, $\tilde{\Gamma}_{ij}$ is positive, leading to $\gamma_{s} > \gamma_{a}$ at $\omega = \omega_{0}$, and so on. Therefore, this switching behavior, from $\gamma_{a} > \gamma_{s}$ to $\gamma_{s} > \gamma_{a}$ and back, follows the same odd $n$-array sequences that are identified in the mediated interactions shown in Fig.~\ref{f2}. 
Thus, depending on the number of particles in the odd-$n$ array, the symmetric state, $|s\rangle$, becomes the subradiant state, while the antisymmetric state, $|a\rangle$, will be superradiant, and vice versa. 
As $n$ increases, Figs. ~\ref{f4} (e)-(h) show that $|\gamma_{a}-\gamma_{s}|$ decreases at $\omega = \omega_{0}$, which is a result of a weakening of the dissipative coupling, since 
$|\gamma_{a}-\gamma_{s}| = 2|\tilde{\Gamma}_{ij}|$. 

In the odd-$n$ array mediators, the qubits are not detuned from the single MNP LSPR ($\Delta_{1} = \Delta_{2} = 0$, so that we have $\Delta_{-} = 0$ again); rather, the mediated entanglement is optimized via the symmetric and antisymmetric decay and excitation rates. 
This allows either the symmetric or antisymmetric state population to contribute most to the degree of entanglement, depending on the odd $n$-array sequence. When $\gamma_{a} > \gamma_{s}$ ($n=3, 7, 11, ...$), we drive the qubits symmetrically ($\Omega_{s} \neq 0, \Omega_{a} = 0$) as shown in 
Fig.~\ref{f3} (b), so that the population of the symmetric state, $|s\rangle$, which is the slowly-decaying state in this case, is optimized over that of $|a\rangle$ for entanglement generation. On the other hand, in Fig. \ref{f3} (c), the qubits are driven antisymmetrically ($\Omega_{s} = 0, \Omega_{a} \neq 0$), since $|a\rangle$ is the subradiant state in this case ($\gamma_{s} > \gamma_{a}$ for $n=5, 9, 13, ...$), allowing a higher population, $\rho_{aa}$, of the antisymmetric state in the degree of entanglement, compared to $\rho_{ss}$. 
Assuming that two laser fields are used to drive the qubits such that $\tilde{\Lambda}_{2} = \tilde{\Lambda}_{1}e^{i\phi}$, with $\phi$ being a phase difference between the two fields, symmetric excitation is realized by a phase difference of $\phi = 0$, while antisymmetric excitation is achieved by setting $\phi = \pi$, resulting in a phase shift of $\pi$ radians between the two laser fields \cite{Martin2011,Li2019}.

\subsection{Stationary concurrence}
We calculate the degree of entanglement generated in the two-qubit system using the concurrence \cite{Wootters1998}
\begin{equation}
	C = \text{max}\Big(0, \lambda_{1} - \sum\limits_{i>1}\lambda_{i}  \Big),
\end{equation}
with $\lambda_{i}$ being the eigenvalues in descending order of the matrix $\sqrt{ \sqrt{\rho}  \tilde{\rho} \sqrt{\rho}}$, with $\tilde{\rho} = (\sigma_{y} \otimes \sigma_{y})\rho^{*}(\sigma_{y} \otimes \sigma_{y})$, where $\rho^{*}$ is the complex conjugate of $\rho$ and $\sigma_{y}$ is the Pauli Y matrix, and $0< C < 1$ for partially entangled states \cite{Tanas2004}.
To obtain the elements of the steady-state density matrix, $\rho^{ss} \equiv \rho(t\rightarrow\infty)$, of the two-qubit system, Eq.~\eqref{e5} was solved in the computational basis at $t\rightarrow\infty$ with both $H$ and $L$ replaced with their effective versions given by Eq.~\eqref{e11a} and Eq.~\eqref{e11b}, respectively. Here, we write the solution in component form, with the two qubits initialized in the ground state, $|0\rangle \langle 0|$, as follows (see Sec.~3 of the SM \cite{Supp} for the full matrix equation):
\begin{equation}
\rho^{ss}_{k} = \sum_{l=1}^{16} (M^{-1})_{kl}\, B_l, \qquad k = 1,2,\dots,16
\end{equation}
where $\rho^{ss}_{k} = \bigl[\rho_{ii}|_{i=0}^{3},\;
\Re\rho_{ij}|_{0\le i<j\le3},\;\Im\rho_{ij}|_{0\le i<j\le3}\bigr]^\top$ and $B_{l} = [1, 0, 0, ..., 0]^\top$ are $16\times1$ column vectors, and $M$ is a $16\times16$ evolution matrix, comprising the effective detuning, damping, coupling, and driving terms of the two-qubit system. 
We have considered $I$ values in the range $0-80$ W/cm$^{2}$, which falls in the weak-excitation regime, $\Omega_{m}/\gamma_{0} \approx 0-0.096$. 

\begin{figure}
\centering 
\includegraphics[width = 0.20\textwidth]{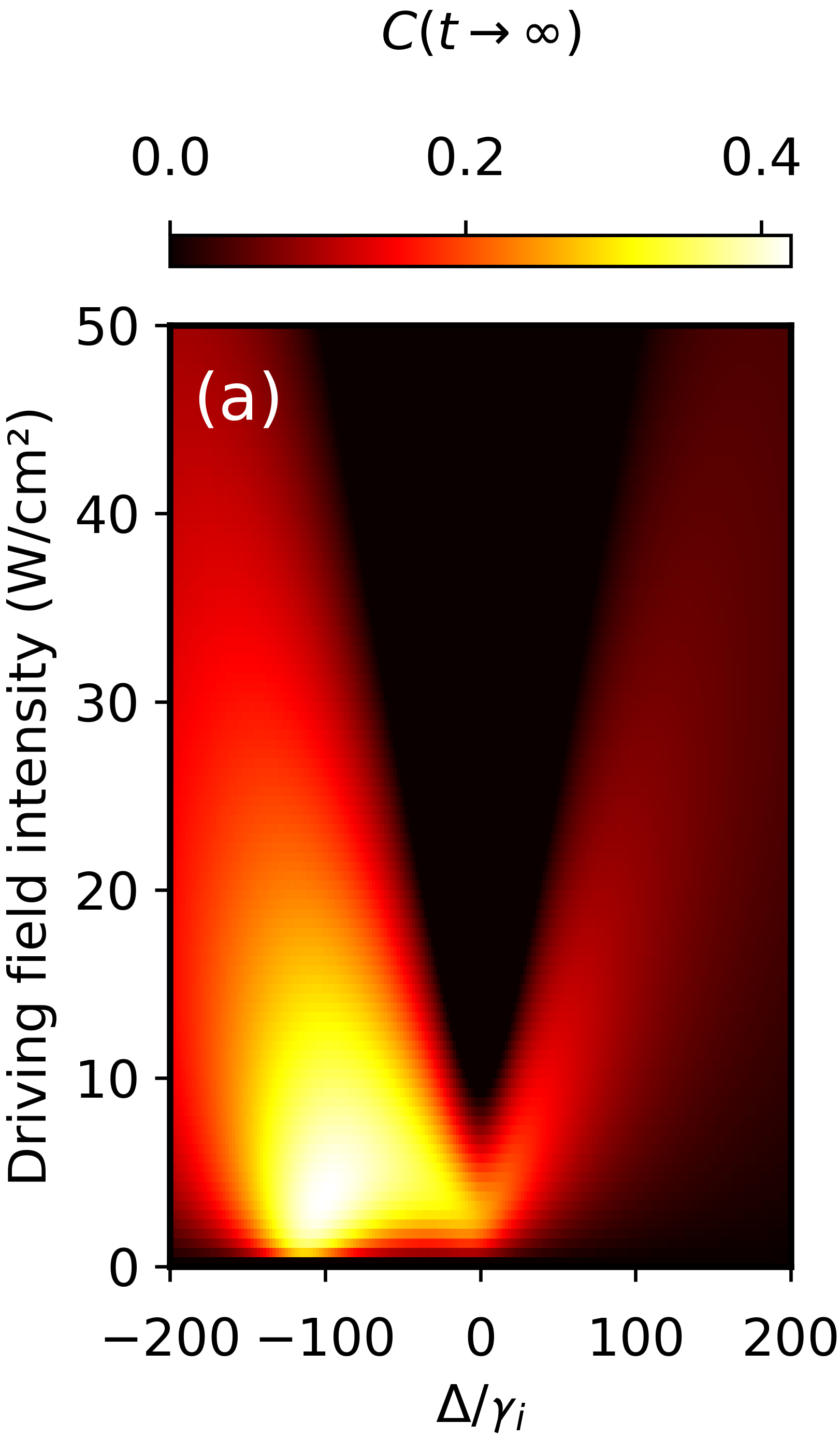}~~~~~
\includegraphics[width = 0.18\textwidth, height=0.258\textheight]{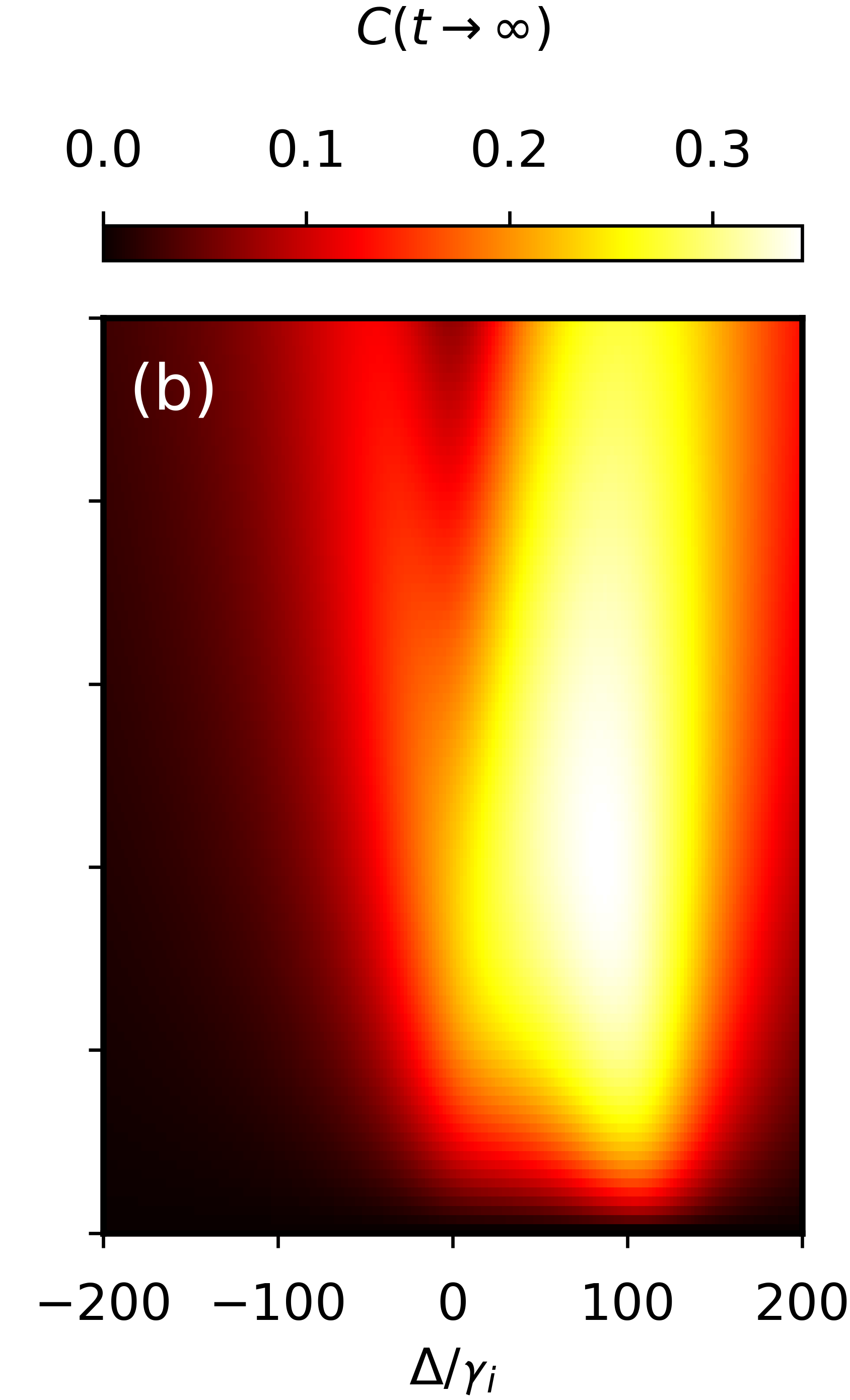} 
\vspace*{0.45cm}\\
\includegraphics[width = 0.20\textwidth]{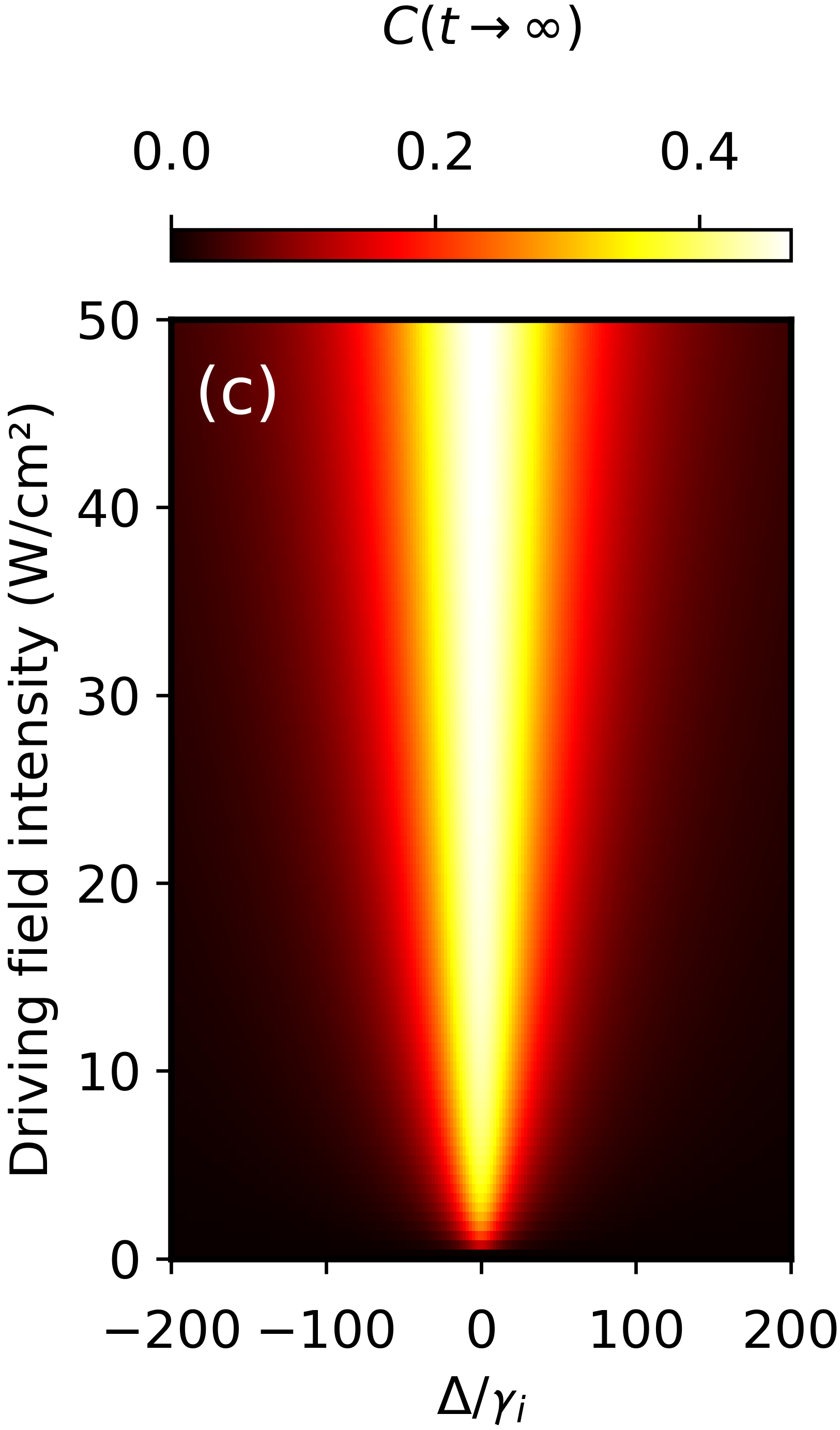}~~~~~
\includegraphics[width = 0.18\textwidth, height=0.258\textheight]{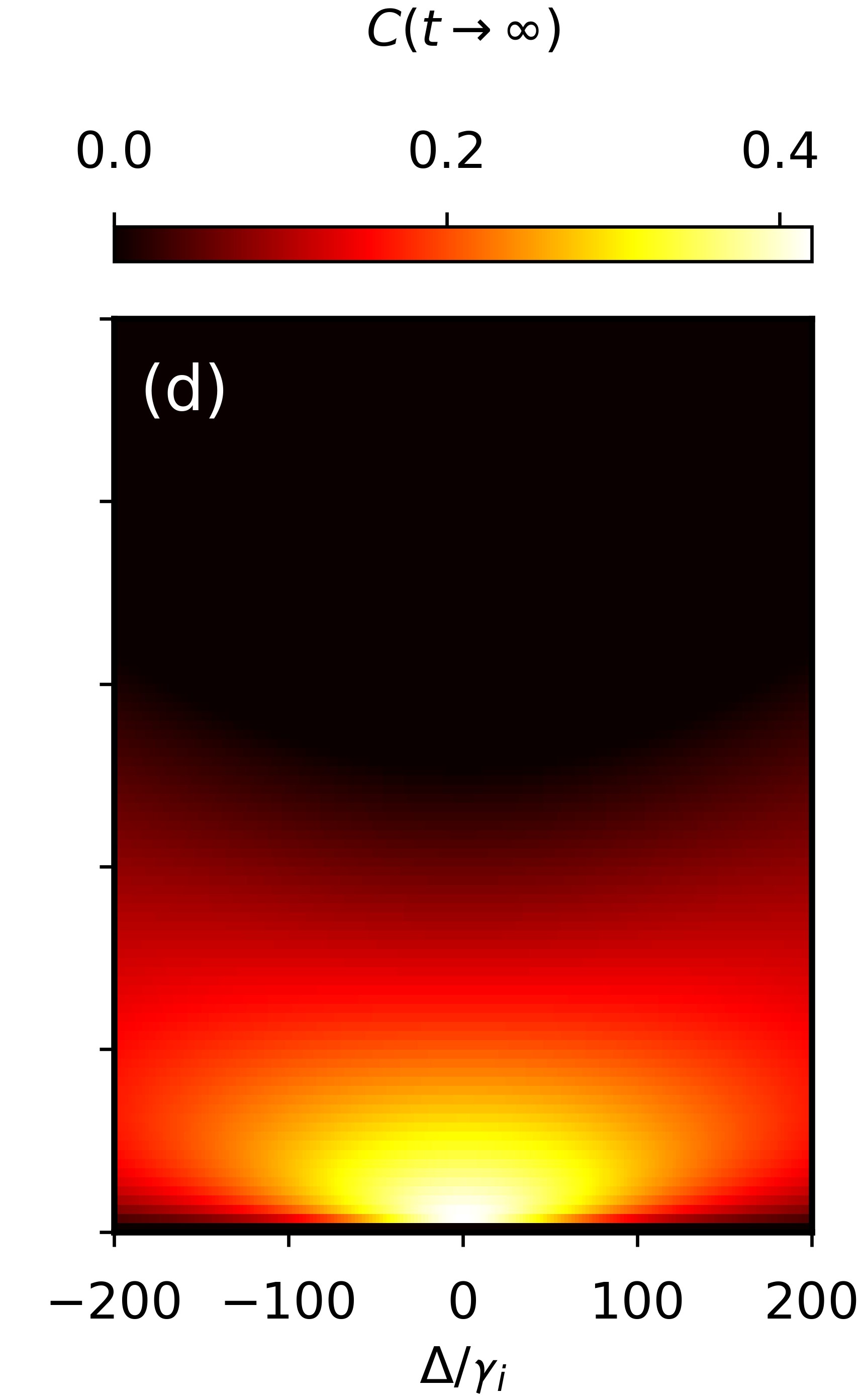}
\caption{(Color online) Dependence of the stationary concurrence on the driving field intensity and the normalized detuning rate, $\Delta/\gamma_{i}$, of each qubit, for the even-$n$ arrays: (a) $n = 2$ and (b) $n = 4$, and the odd-$n$ arrays: (c) $n = 3$ and (d) $n = 5$.
}\label{f8}
\end{figure}

In order to determine the detuning rate, $\Delta$, of each qubit needed to optimize the mediated entanglement with the even-$n$ arrays, we investigated the dependence of the stationary concurrence, $C(t\rightarrow\infty)$, on values of $\Delta$ ranging from $-200\gamma_{i}$ to $200\gamma_{i}$ for the range of driving field intensities given above, as shown in Fig.~\ref{f8}. Fig.~\ref{f8} (a) shows that $C(t\rightarrow\infty)$ is maximized for $n=2$ when $\Delta \sim -70\gamma_{i}$ to $-100\gamma_{i}$, i.e., negative detunings. This is the case for the even-$n$ array sequence $n = 2, 6, 10, ...$. 
On the other hand, Fig.~\ref{f8} (b) shows that $C(t\rightarrow\infty)$ is maximized for $n=4$ when $\Delta \sim 70\gamma_{i}$ to $100\gamma_{i}$, i.e., positive detunings. 
This is also the case for the even-$n$ array sequence $n = 4, 8, 12, ...$.
However, when $n$ is odd, Fig.~\ref{f8} (c) and (d) show that the concurrence is maximized when $\Delta = 0$, since odd-$n$ arrays support a hybrid LSPR at the driving frequency: $\omega = \omega_{0} = \omega_{i}$. 
Based on the stationary populations of the Dicke states in our model,
shown in Fig.~S3 of the SM \cite{Supp}, 
we observed that these values of $\Delta$ optimize $\rho_{ss}$ over $\rho_{aa}$, $\rho_{ee}$, and $\rho_{gg}$, for entanglement generation mediated by even-$n$ arrays. 
We propose that this is because these values of $\Delta$ are within the same order of magnitude as the coherent coupling rate ($\sim 10^{10}$~rad/s) shown in Fig.~\ref{f2} (a). Therefore, we only consider qubit detunings, $\Delta$, for which $|\Delta| \approx |\tilde{g}|$. 
For the even-$n$ array sequence $n = 2, 6, 10, ...$, some negative detunings in Fig.~\ref{f8} (a) satisfy this condition, since $\tilde{g} = \tilde{G}_{ij} < 0 $, as shown in Fig.~\ref{f2} (a), while for the even-$n$ array sequence $n = 4, 8, 12, ...$, the condition is met via some positive detunings in Fig.~\ref{f8} (b), since $\tilde{g} = \tilde{G}_{ij} > 0$, as shown in Fig.~\ref{f2} (a). 

\begin{figure} 
	\centering 
	\includegraphics[width=0.4\textwidth]{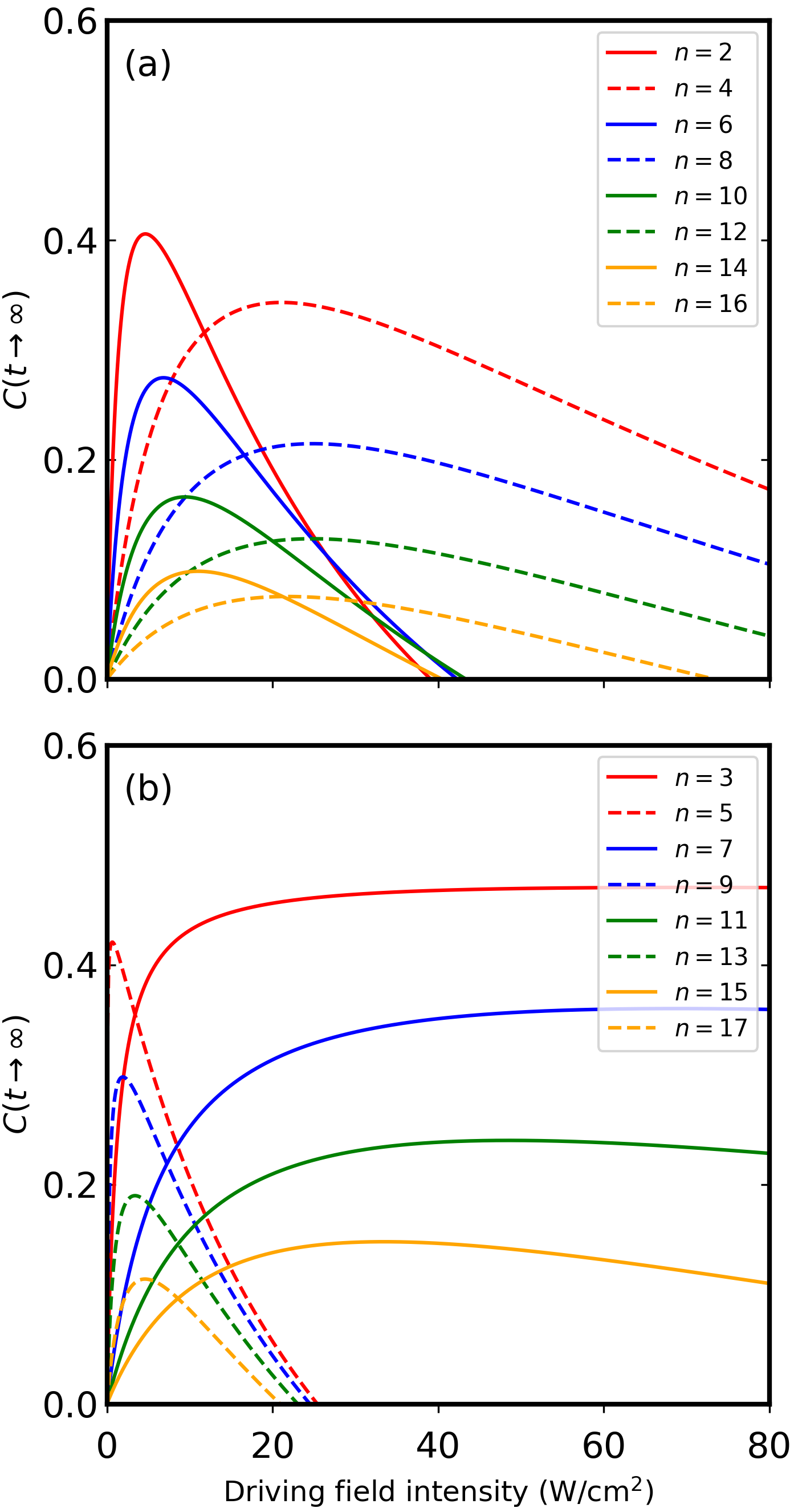}
	\caption{(Color online) Dependence of the stationary concurrence on the intensity of the driving field for (a) even-$n$ arrays $n=2, 6, 10, 14$ (solid curves) and $n= 4, 8, 12,16$ (dashed curves), and (b) odd-$n$ arrays $n=3, 7, 11, 15$ (solid curves) and $n= 5, 9, 13, 17$ (dashed curves). For each $n$-array mediator, the concurrence has been optimized based on the energy-level diagrams in Fig.~\ref{f3}. 
	}\label{f5}
\end{figure}

Fig.~\ref{f5} shows the dependence of the stationary concurrence on the driving field intensity. 
We have set $\Delta = \pm 80\gamma_{i}$ ~($\sim \pm 4.4\times10^{10}$~rad/s), which is within the optimal range of $\Delta$ values for the stationary concurrence shown in Fig.~\ref{f8}. 
As shown in Fig.~\ref{f5} (a), the concurrence for each even-$n$ array reaches a maximum value at a certain driving field intensity, beyond which it decreases with increasing intensity. Since $C \approx \text{max}(0, \sqrt{(\rho_{ss}-\rho_{aa})^{2} +4\Im\rho_{sa}^{2}} - 2\sqrt{\rho_{gg}\rho_{ee}})$, with $\rho_{ee}, \rho_{gg}, \rho_{ss}, \rho_{aa}$, and $\rho_{sa}$ being the populations of $|e\rangle, |g\rangle, |s\rangle$, $|a\rangle$, and the coherence between $|s\rangle$ and $|a\rangle$, respectively \cite{Tanas2004,Hou2014}, we observed from our model that the  decrease in concurrence with an increase in the field intensity is due to an increase in the population, $\rho_{ee}$, of the bi-excited state, and a decrease in the population, $\rho_{ss}$, of the symmetric state, while $\rho_{aa} = \rho_{ee}$ and $\Im \rho_{sa} = 0$ (see Fig.~S3 of the SM \cite{Supp}). 

When the qubit--qubit interactions are mediated by the even $n$-array sequence $n = 2, 6, 10, ...$ (solid curves in Fig.~\ref{f5} (a)), the optimal concurrence is attained at weaker driving fields ($I\propto n$) compared to mediation with the even $n$-array sequence $n = 4, 8, 12, ...$ (dashed curves in Fig.~\ref{f5} (a)), where the optimal driving field is $\gtrsim 20$ W/cm$^{2}$.
We attribute this to the behavior of $|\Omega_{s}|$ and $\gamma_{s} = \tilde{\gamma}_{i}$ at $\omega = \omega_{0}$. For instance, $|\Omega_{s}| (n=2) > |\Omega_{s}| (n=4)$, as shown in Fig.~S5 of the SM \cite{Supp}, and $\gamma_{s} (n=4) > \gamma_{s} (n=2)$ as we have shown in Fig.~\ref{f4} (a), so that a higher intensity is required to overcome both the weaker excitation rate and higher damping, and attain an optimal concurrence with $n=4$. This trend re-occurs in the even $n$-array pair $n = 6, 8$, and so on. 
As $n$ increases in each even $n$-array sequence, the symmetric decay rate, $\gamma_{s} = \gamma_{a} = \tilde{\gamma}$, is enhanced via an increase in the spontaneous emission rate, $\tilde{\gamma}_{i}$, of each qubit, as shown in Figs.~\ref{f4} (a)-(d), while the energy gap $|E_{+} - E_{-}| = 2|\tilde{G}_{ij}|\rightarrow 0$ at $\omega = \omega_{0}$. As shown in Fig.~S3 of the SM \cite{Supp}, an increase in $n$ leads to higher values of $\rho_{gg}$ and lower values of $\rho_{ee}$, $\rho_{ss}$, and $\rho_{aa}$, while $\Im\rho_{sa} = 0$. As a result, the concurrence decreases. 

Fig.~\ref{f5} (b) shows that the degree of stationary qubit--qubit entanglement is not only preserved but also slightly enhanced in longer NAs via odd $n$-array mediators compared to even ones. 
The enhancement comes from either zero values of $\rho_{ee}$ or slightly lower values of $\rho_{ee}$ or $\rho_{gg}$ in comparison to the even $n$-array mediators, depending on the odd $n$-array sequence, as shown in Figs.~S3 and S4 of the SM \cite{Supp}. 
The optimal concurrence at $\omega = \omega_{0}$ follows the two sequences previously identified in the dissipative coupling shown in Fig.~\ref{f2}. 
With the NA sequence $n = 3, 7, 11, ...$ (solid curves in Fig.~\ref{f5} (b)), the concurrence reaches an optimal value at a certain intensity ($40 \lesssim I\lesssim60$ W/cm$^{2}$), beyond which it increases only slightly ($n = 3$ and $n = 7$) or decreases ($n = 11$ and $n = 15$) with increasing field intensity 
between $56$ W/cm$^{2}$ and $80$ W/cm$^{2}$. A further increase in the driving field intensity beyond $100$ W/cm$^{2}$ (see Fig.~S6 of the SM \cite{Supp}) eventually leads to a decrease in the concurrence for $n=3$ and $n = 7$, due to an increase in $\rho_{ee}$, but this is already approaching the strong driving regime, where the validity of the effective model remains questionable \cite{Hou2014}. 

However, with the NA sequence $n = 5, 9, 13, ...$ (dashed curves in Fig.~\ref{f5} (b)), $\rho_{aa}$ contributes most to the degree of entanglement (see Figs.~S4 of the SM \cite{Supp}), 
due to dissipative coupling to $|a\rangle$ via $\gamma_{a} > \gamma_{s}$, combined with $\Omega_{s} = 0, \Omega_{a} \neq 0$, but 
there are non-zero values of other populations contributing to the entanglement. 
Hence, in this case, the concurrence peaks at smaller field intensities ($\lesssim3$ W/cm$^{2}$), as shown in Fig.~\ref{f5} (b) (dashed curves) and in Fig.~\ref{f8} (d) for $n=5$. 
The weaker driving fields required to reach the optimal concurrence with the NA sequence $n = 5, 9, 13, ...$ in comparison to the sequence $n = 3, 7, 11, ...$, in Fig.~\ref{f5} (b), can be explained based on the behavior of $|\Omega_{s}|$ and $ |\Omega_{a}|$, as well as $\gamma_{s}$ and $\gamma_{a}$. 
For instance, at $\omega = \omega_{0}$, $|\Omega_{s}| (n = 3) << |\Omega_{a}| (n = 5)$ (see Fig.~S5 of the SM \cite{Supp}) and $\gamma_{a} (n = 3) > \gamma_{s} (n = 5)$ as shown in Fig.~\ref{f4}, so that a higher field intensity is required to overcome the increased damping and reach the optimal concurrence when $n = 3$. This repeats for the next odd-$n$ pair $n = 7, 9$, and so on. 

The enhanced excitation rates of the qubits when $n = 5, 9, 13, ...$ are due to the presence of a hybrid plasmon mode with the same frequency as the single MNP LSPR in the excitation spectrum (Fig.~S5 of the SM \cite{Supp}) at $\omega = \omega_{0}$, i.e.,
when coupled to the the NA sequence $n = 5, 9, 13, ...$, the qubits experience a resonant excitation while with the sequence $n = 3, 7, 11, ...$, they undergo an off-resonant excitation (see Fig.~S5 of the SM \cite{Supp}). 
As $n$ increases, the decay splitting $|\gamma_{a}- \gamma_{s}| = 2|\tilde{\Gamma}_{ij}| \rightarrow 0$, as shown in Figs. \ref{f4} (e)-(h), in each odd $n$-array sequence. 
The dependence of the stationary populations and coherence on the driving field intensity and number of NA mediators is shown in Fig.~S4 of the SM \cite{Supp}. For both NA sequences ($n = 3, 7, 11,...$, top row and $n = 5, 9, 13,...$, bottom row), $\rho_{gg}$ decreases and $\rho_{ee}$ increases with an increase in $n$ and $\Im\rho_{sa} = 0$, but $\rho_{aa}$ and $\rho_{ss}$ show substantial differences between the two sequences, with $\rho_{aa} = \rho_{ee}$ and $\rho_{ss}$ decreasing with increasing $n$ for the sequence $n = 3, 7, 11,...$, while $\rho_{aa}$ decreases with increasing $n$ and $\rho_{ss} = \rho_{ee}$ for the sequence $n = 5, 9, 13, ...$.
 causing a decrease in the concurrence. The behavior of the concurrence due to the even and odd $n$-arrays is therefore similar to those of even and odd cavity modes in two-qubit entanglement generation in the presence of a plasmonic nanocavity \cite{Crookes2025}.
 
\subsection{Entanglement decay}
\begin{figure} 
	\centering 
	\includegraphics[width=0.4\textwidth]{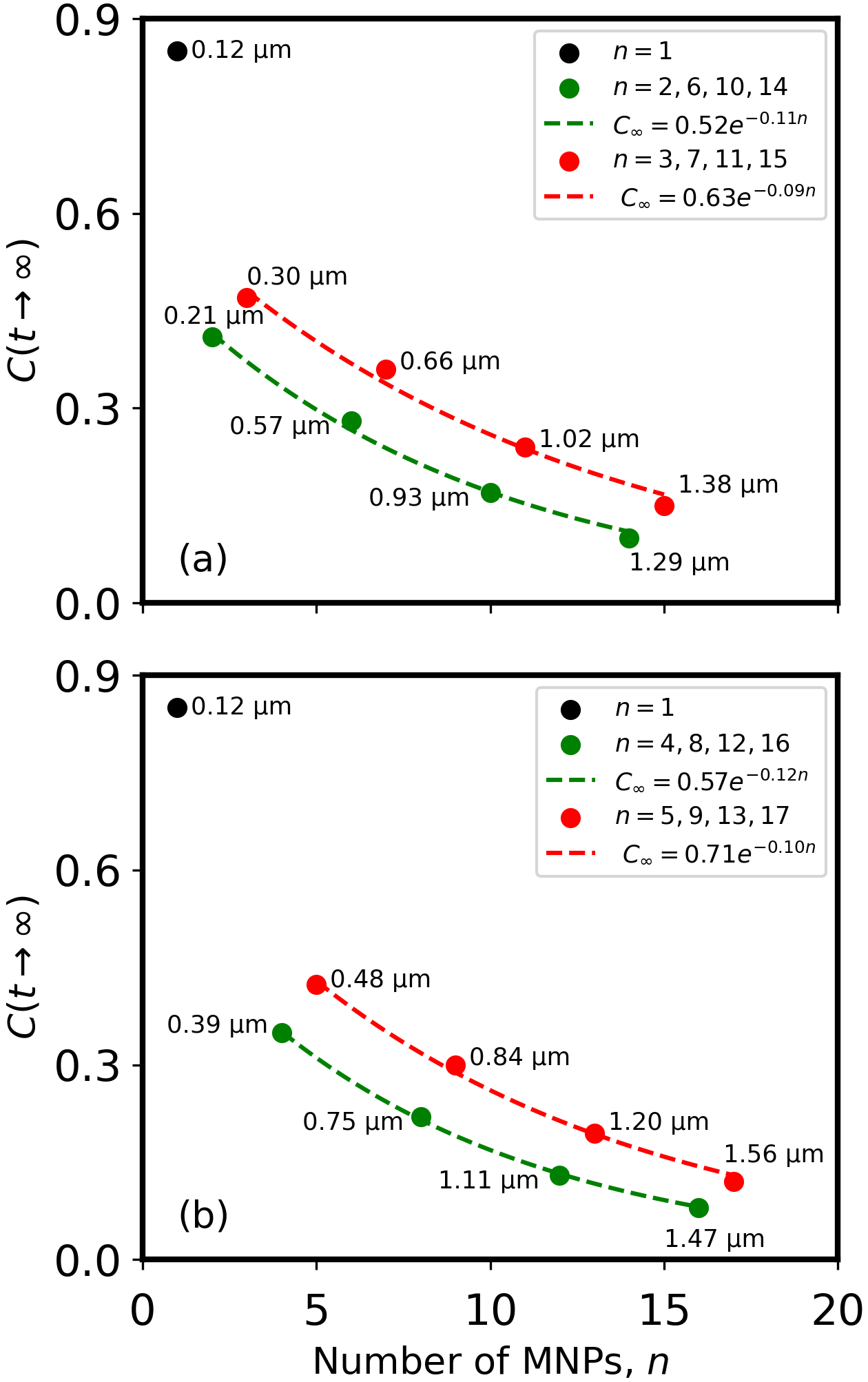}
	\caption{(Color online) Decay of stationary concurrence with increase in the number, $n$, of MNPs in the array. (a) NA sequences: $n = 2, 6, 10, 14$ and $n = 3, 7, 11, 15$. (b) NA sequences: $n = 4, 8, 12, 16$ and $n = 5, 9, 13, 17$. The inter-qubit distance, corresponding to the number of MNPs in each NA via $d_{qq}-2r_{0}=2rn + s(n+1)$, is indicated in $\mathrm{\mu}$m. 
		Solid points: data. Dashed lines: decay fits. 
	}\label{f6}
\end{figure}

We now compare the stationary concurrence for even and odd $n$-array mediators, obtained at $\omega = \omega_{0}$, based on the optimal driving field intensities discussed in the previous section.  
Fig.~\ref{f6} shows the dependence of the concurrence on the number of MNPs, $n$, in the NA. To serve as a reference, we have included the optimal stationary concurrence obtained for $n=1$, based on antisymmetric detuning of the qubits, as has been extensively studied in previous works \cite{He2012, Hou2014,Nerk2015,Dumit2017}. 
When $n=1$, the concurrence is very high ($C_{\infty} \approx 0.85$), but the inter-qubit distance is very short ($d_{qq}-2r_{0} = 0.12~\mathrm{\mu}$m), as shown in Fig.~\ref{f6}. However, a single MNP is inefficient in long-range entanglement mediation because the initially high concurrence rapidly decays with an increase in the MNP-QD separation. 
The fitted curves---dashed lines in Fig.~\ref{f6}---show that the stationary concurrence follows a decay law of the form $C_{\infty} = C_{0}e^{-\tau n}$, where $\tau$ is the decay constant of stationary concurrence per MNP in the array.  
In the limit of very long arrays ($n \rightarrow \infty$), we find that the entanglement vanishes, since $C_{\infty} \rightarrow 0$. 

Fig.~\ref{f6} (a) shows that the odd $n$-array mediators (red curve) not only sustain the degree of entanglement at longer inter-qubit distances---$90$ nm greater than those of a neighboring even-$n$ array---but also support a slightly higher degree of entanglement. This trend persists even in longer arrays, as shown in Fig.~\ref{f6} (b). We attribute the robustness of the odd $n$-array mediators to entanglement decay ($C^{odd~n}_{0} > C^{even~n}_{0}$  and $\tau_{odd~n} < \tau_{even~n}$), to the enhanced dissipative coupling in comparison to the coherent coupling ($\tilde{\Gamma}_{ij} \approx 10~\tilde{G}_{ij}$ at the driving frequency $\omega = \omega_{0}$, as shown in Fig.~\ref{f2}). For this reason, the decay splitting $|\gamma_{a}-\gamma_{s}| = 2|\tilde{\Gamma}_{ij}|$ for odd $n$ is large (Fig.~\ref{f4} (e)-(h)), and therefore more effective than the energy splitting $|E_{a}-E_{s}| = 2|\tilde{G}_{ij}|$ for even $n$ in trapping a higher population of the NA-coupled qubits in either the symmetric or antisymmetric state. 
Hence, with NAs, the inter-particle coupling, $\kappa$, enables the concurrence to be sustained over long inter-qubit distances ($\mathrm{\mu}$m regime), in contrast to a single MNP where the entanglement decay with an increase in the inter-qubit separation is much more rapid. 

\subsection{Model validation}
We validated the effective model of the NA-mediated stationary entanglement based on simulations performed using the Quantum Optics Toolbox in Python (QuTiP) \cite{Johan2012}. 
The QuTiP simulations were set up using the total Hamiltonian and Liouvillian given by Eqs.~\eqref{e2} and \eqref{e4}, respectively, to obtain the steady-state density matrix of the coupled system. 
The concurrence was then calculated using a density matrix of the qubits---a reduced density matrix obtained by performing a partial trace on the steady-state density matrix. 
\begin{figure} 
	\centering 
	\includegraphics[width=0.4\textwidth]{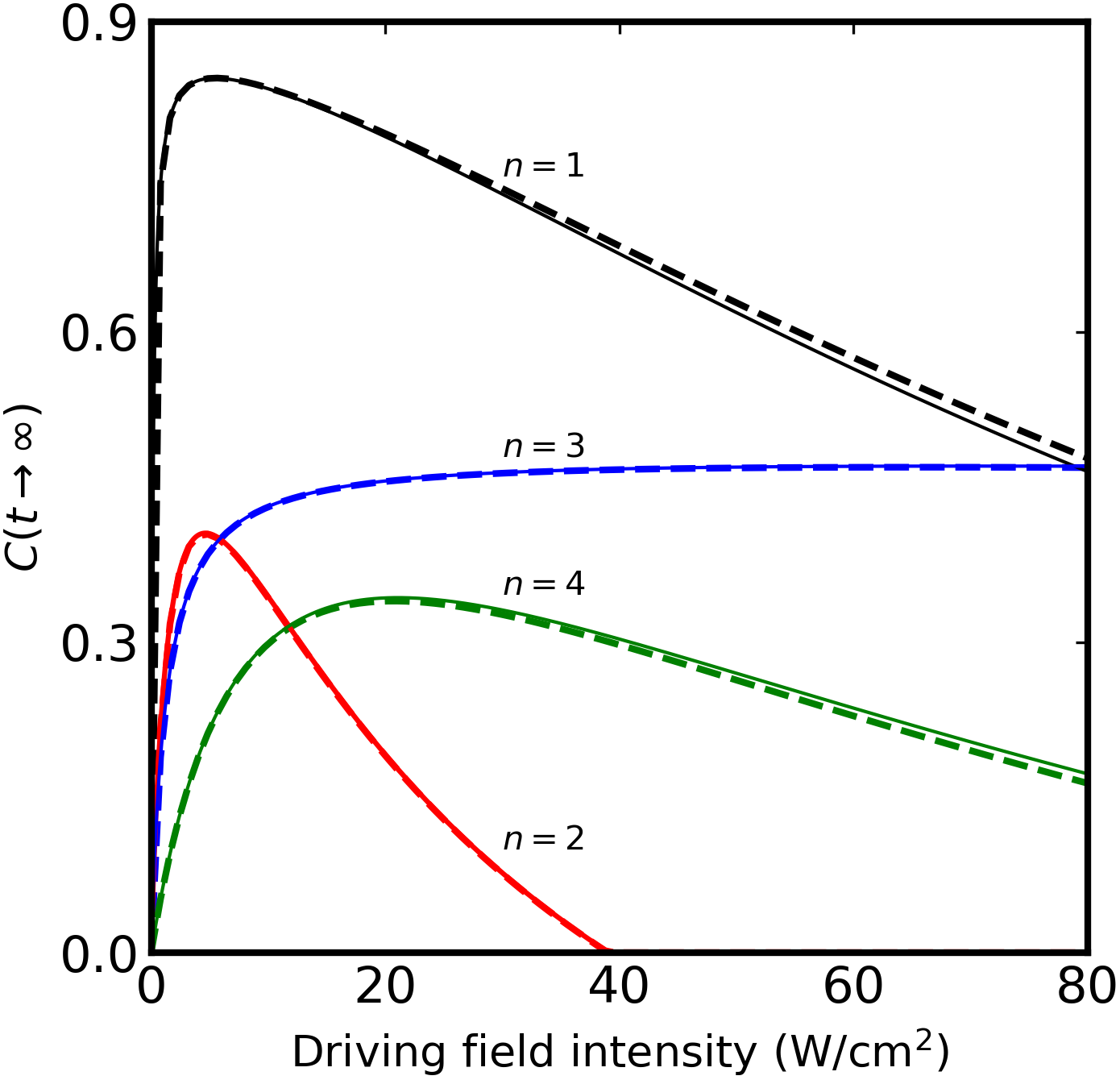}
	\caption{(Color online) Effective model (solid curves) versus simulation (dashed curves) of the mediated stationary concurrence for $n = $1, 2, 3, and 4. 
	}\label{f7}
\end{figure}
Fig.~\ref{f7} shows a comparison between the stationary concurrence simulated in QuTiP and calculated analytically using the effective model. There is good agreement between the simulation and the effective model for $n = $1, 2, 3, and 4. 

Here, we have only shown the simulated stationary concurrence up to $n=4$, but the concurrence due to higher values of $n$ can also be simulated. However, the dimension of the Hilbert space required to perform the simulation is proportional to $2^{2} \text{(QD qubits)} \times N^{n} \text{(MNPs)}$, where each MNP is modeled as a quantum harmonic oscillator (QHO) with levels $N \ge 2$, representing at least the ground and first excited state of the QHO. To improve the accuracy of the simulated results shown in Fig.~\ref{f7}, we have used $N = 4$. 
Hence, as $n$ increases, more computational time and memory are needed to simulate the concurrence compared to the effective approach. 
The effective model, therefore, offers additional benefits beyond providing an in-depth theoretical insight into the behavior of the mediated interactions. 

\section{Conclusion}\label{section5}
We have investigated the plasmonic mediation of stationary entanglement in two QD qubits using an effective cavity quantum electrodynamic approach.
This has allowed us to obtain a comprehensive insight into the role of plasmon-mediated interactions in long-range entanglement mediation with one-dimensional plasmonic nanoarrays. 
Within the weak-coupling and weak-excitation limits, the effective model has enabled us to predict MNP number-dependent mediated interactions---coherent and dissipative coupling rates---due to even and odd $n$-array mediators of the two-qubit entanglement.  
We found that with even-$n$ arrays, the mediated entanglement is due to the coherent coupling $\tilde{G}_{ij}$, while the dissipative coupling $\tilde{\Gamma}_{ij}$ is responsible for the degree of entanglement obtained with odd-$n$ arrays. 
This enables the mediated qubit--qubit entanglement to be optimized via energy splitting ($|E_{a}-E_{s}| = 2|\tilde{G}_{ij}|$) when $n$ is even or via decay splitting ($|\gamma_{a}-\gamma_{s}| = 2|\tilde{\Gamma}_{ij}|$) when $n$ is odd. 
Through resonant driving at the single MNP LSPR, the dissipative coupling undergoes a resonant enhancement that enables odd-$n$ arrays to outperform even-$n$ arrays as entanglement mediators. 

One might expect the mediated entanglement to decay straightforwardly with increasing number of MNPs in the array but this is not the case. Rather, the entanglement decay inherits the same MNP array sequences found in the mediated interactions. 
Although the nanoarrays suffer from high ohmic losses and increased weakening of the mediated interactions, which subsequently lead to entanglement decay, we have shown that the proposed scheme is capable of sustaining non-vanishing stationary two-qubit entanglement beyond a qubit--qubit distance of one micron. 
This decaying entanglement can be distilled \cite{Ben1996,Zhao2003} or stabilized \cite{Irfan2024,Vivas2024} for use in a quantum information protocol, such as quantum sensing \cite{Hyllus2012,Toth2020,Trenyi2024,Demille2024,Lee2021}. 

\section*{Acknowledgements}
This research was financially supported by the Department of Science, Technology and Innovation (DSTI) through the South African Quantum Technology Initiative (SA QuTI), Stellenbosch University (SU), the National Research Foundation (NRF), and the Council for Scientific and Industrial Research (CSIR).

\bibliographystyle{unsrt}
\bibliography{manuscript.bib}
\end{document}